\documentclass[letterpaper, 10 pt, conference]{ieeeconf}  

\IEEEoverridecommandlockouts         

\usepackage{amsmath}  
\usepackage{amssymb}  
\usepackage{graphicx}
\usepackage{bm}       
\usepackage{xcolor}
\newtheorem{assumption}{Assumption}
\newtheorem{theorem}{Theorem}

\newtheorem{lemma}{Lemma}

\newtheorem{definition}{Definition}

\title{\LARGE {\bf
Linear quadratic regulation of polytopic time-inhomogeneous \\ 
Markov jump linear systems (extended version)
}}
\author{Yuriy Zacchia Lun, 
Alessandro Abate and
Alessandro D'Innocenzo 
\thanks{Y. Zacchia Lun is with IMT School for Advanced Studies Lucca, Italy. 
A. D'Innocenzo is with the Center of Excellence DEWS 
and with the Department of Information Engineering, 
Computer Science and Mathematics of the University of L'Aquila, Italy. 
A. Abate is with the 
Department of Computer Science of the University of Oxford, UK.
The research leading to these results has received funding from the Italian 
Government under CIPE resolution n.135 (Dec. 21, 2012), project \emph{INnovating 
City Planning through Information and Communication Technologies 
(INCIPICT).}
}
}

\begin{document}

\maketitle
\thispagestyle{empty}
\pagestyle{empty}

\begin{abstract}
In most real cases transition probabilities between 
operational modes of Markov jump linear systems cannot be computed 
exactly and are time-varying. We take into account this aspect by 
considering Markov jump linear systems where the underlying 
Markov chain is polytopic and time-inhomogeneous, i.e.~its 
transition probability matrix is varying over time, with variations that are 
arbitrary within a polytopic set of stochastic matrices.
We address and solve for this class of systems the infinite-horizon optimal 
control problem. In particular, we show that the optimal controller can be 
obtained from a set of coupled algebraic Riccati equations,
and that for mean square stabilizable systems the optimal 
finite-horizon cost corresponding to the solution to a 
parsimonious set of coupled difference Riccati equations converges 
exponentially fast to the optimal infinite-horizon cost related to the set 
of coupled algebraic Riccati equations. 
All the presented concepts are illustrated on a numerical example 
showing the efficiency of the provided solution.
\end{abstract}

\section{INTRODUCTION}
In discrete-time Markov jump linear systems (MJLSs), the transition probabilities 
of jumps between operational modes are fundamental in determining the dynamic 
behaviour \cite{costa2006discrete}. These transition probabilities are generally 
considered to be either time-invariant or certain in the majority of dedicated 
studies, see \cite{zhang2016analysis} as a textbook presenting important 
results and detailed examination of the general state of the art.

In most real cases, however, the transition probability matrices (TPMs) are 
affected by global uncertainty due to random and systematic errors of measurement 
and numerical computation procedures (used to obtain the values of TPMs), by 
incomplete knowledge of some transition probabilities (when adequate samples of 
the transitions are costly or time-consuming to obtain), and by abrupt and 
unpredictable time-variance (due to environmental factors, like for instance the 
wind perturbing the model of airspeed variation in a vertical take-off landing 
helicopter system \cite{long2013fault}). The study of robustness to such variations 
is naturally important in many applications, especially for wireless networked
control systems (see e.g.~\cite{Hespanha2007,4118476,PajicTAC2011,AlurTAC11} 
and references therein for a general overview)

In this work, we allow for incomplete knowledge and time-varying uncertainties 
in transition probabilities by studying polytopic time-inhomogeneous (PTI) MJLSs,
where the TPMs are unknown and time-varying within a bounded set.
This model permits to include realistic and reasonable uncertainties while still 
maintaining the problem convex. In fact, there exists a considerable number 
of works on discrete-time Markov jump systems (both linear and nonlinear) with 
polytopic uncertainties, which can be either time-varying or time-invariant, as 
extensively discussed in \cite[Section 1.5]{yzl_dissertation}.

We build upon our previous results on stability \cite{yzl2016cdc} and optimal 
finite-horizon control \cite{8264642} of PTI MJLSs, providing an 
analytical solution for the infinite-horizon optimal state-feedback control problem, 
that can be computed efficiently. Specifically, we show that for a stabilizable 
system, the optimal steady-state controller is obtained from a set of coupled 
algebraic Riccati equations (CAREs). The cardinality of this set of CAREs equals 
to the number of vertices of the convex polytope characterizing the time-varying 
uncertainties in TPMs. Furthermore, when a PTI MJLS is stabilizable (in the mean 
square sense), the optimal finite-horizon cost of the robust control obtained 
from the solution to a parsimonious set of coupled difference Riccati equations 
(CDREs) converges exponentially fast to the optimal infinite-horizon cost related
to the set of CAREs.
From a technical point of view these results are a nontrivial 
extension of \cite{costa2006discrete,yzl2016cdc} and \cite{8264642}, since they
require a proper definition of an appropriate set of CAREs, which solution exists, 
is unique, and achieves the optimal quadratic cost, while the convergence between 
the finite- and infinite-horizon controllers is ensured in terms of costs of 
control actions and not in terms of the Riccati equations themselves. The obtained
results are validated on a numerical example based on the Samuelson's 
multiplier-accelerator model with 
three operational modes, three to four polytopic bounds on a time-inhomogeneous 
TPM and different time horizons, showing that for a stabilizable 
and detectable system both finite- and infinite-horizon control problems can be 
solved efficiently by taking advantage of parsimonious sets.

This paper is organized as follows. After presenting the PTI model of MJLSs,
in Section \ref{sec:pb_statement}, we summarize our result on finite-horizon
optimal robust state-feedback control in a way useful to formulate and solve 
the problem for the infinite-time horizon. In Section \ref{sec:mss}, we 
recall the concept of the stability equivalence in PTI setting and formally 
define the notion of mean square stabilizability 
and mean square detectability.
In Section \ref{sec:care}, we define the stabilizing solution to the control 
CAREs, show that, for mean square stabilizable 
and detectable PTI MJLSs, this solution is unique, 
it achieves the optimal infinite-horizon cost of robust state-feedback control, 
and the optimal finite-horizon cost converges to it. 
In Section \ref{sec:example},
we showcase our results on a numerical example.
Lastly, Section \ref{sec:conclusions} concludes the paper.

\subsection{Notation}
We denote the set of all either real or complex numbers by $\mathbb{F}$, and the 
sets of integers, of all nonnegative integers, and of all positive integers by 
$\mathbb{Z}$, $\mathbb{Z}_0$, and $\mathbb{Z}_+$, respectively. An $n$-dimensional 
linear space with entries in $\mathbb{F}$ is indicated by $\mathbb{F}^{n}$, while 
a set of matrices with $m$ rows, $n$ columns, and entries in $\mathbb{F}$ is 
denoted by $\mathbb{F}^{m,n}$. The sets of positive definite and positive 
semi-definite matrices of order $n$ are indicated by $\mathbb{F}^{n,n}_+$ and 
$\mathbb{F}^{n,n}_0$, respectively. As alternative notation to 
$A \!\in\! \mathbb{F}_{0}^{n,n}$ (respectively to $A \!\in\! \mathbb{F}_{+}^{n,n}$) 
we also write $A \!\succeq\! 0$ (respectively $A \!\succ\! 0$). The identity 
matrix of size $n$ is denoted by $I_n$.
The operation of transposition is indicated by superscript $^{T}$, the complex 
conjugation by overbar, while the conjugate transpose of a (complex) matrix is 
denoted by superscript $^*$. Clearly, for a set of real matrices, the transpose 
and conjugate transpose are the same. Then, $\otimes$ denotes the Kronecker product,
and $\oplus$ stands for the direct sum. Notably, the direct sum of a sequence of 
square matrices ${\bm A} \!=\! \left( A_i \right)_{i = 1}^{N}$ produces a block 
diagonal matrix, having the elements of ${\bm A}$ on the main diagonal blocks. 
The spectral radius of a square matrix is indicated by $\rho(\cdot)$, while the 
joint spectral radius of a set of square matrices is denoted by $\hat{\rho}(\cdot)$. 
Finally, $\|\cdot\|$ stands for (induced) Euclidean norm, 
$\mathrm{conv}$ indicates the convex hull of a nonempty set,
$\mathbb{E}(\cdot)$ denotes the expected value.

\section{PROBLEM STATEMENT}\label{sec:pb_statement}
Consider the stochastic basis 
$\left(\Omega, \mathcal{F}, \left( \mathcal{F}_k \right), \mathrm{Pr} \right)$,
where 
$\Omega$ is the sample space, 
$\mathcal{F}$ is the $\sigma$-algebra of (Borel) measurable events,
$\left( \mathcal{F}_k \right)$ is the related filtration, and 
$\mathrm{Pr}$ is the probability measure.
An 
MJLS 
defined on this stochastic basis is
represented by the following dynamical system

\vspace*{-2mm}
\begin{small}
\begin{equation}\label{eq:model_mjls_control}
\begin{cases}
\mathrm{x}_{k+1}=A_{\theta_k} \mathrm{x}_k\!+\!B_{\theta_k} \mathrm{u}_k, \\
\quad \mathrm{z}_k=C_{\theta_k} \mathrm{x}_k\!+\!D_{\theta_k} \mathrm{u}_k, 
\end{cases}
\end{equation}
\end{small}

\vspace*{-3mm}
\noindent where
$\mathrm{x}_k \!\in\!\mathbb{F}^{n_{\mathrm{x}}}$ is the state vector,
$k \!\in\!\mathbb{Z}_0$ is a discrete-time instant,
$\mathrm{u}_k \!\in\! \mathbb{F}^{n_{\mathrm{u}}}$ is the controlled input, and
$\mathrm{z}_k \!\in\! \mathbb{F}^{n_{\mathrm{z}}}$ is the vector of measured 
system output variables.
Then, $\theta \!:\! \mathbb{Z}_0 \!\times\! \Omega \!\to\! \mathbb{M}$ 
is a discrete-time Markov chain, that takes values in a finite set of operational 
modes $\mathbb{M} \!\triangleq\! \{ i \}_{i=1}^N$, so for every operational mode 
there is a correspondent system matrix of appropriate size, and the collection 
of the system matrices of each type is represented by a sequence of $N$ matrices.
Specifically, ${\bm A} \!\triangleq\!( A_i )_{i=1}^{N}$
is the sequence of state matrices, ${\bm B} \!\triangleq\! ( B_i )_{i=1}^{N}$ 
is the sequence of input matrices,
${\bm C} \!\triangleq\! ( C_i )_{i=1}^{N}$
is the sequence of output matrices, while 
${\bm D} \!\triangleq\! ( D_i )_{i=1}^{N}$
is the sequence of direct transition matrices. 
Finally, the initial state and the initial operational 
mode are $\mathrm{x}_0$ and $\theta_0$, respectively. For ease of notation, 
from here on, we denote the initial condition as 
$\phi\!\triangleq\!(\mathrm{x}_0, \theta_0)$, and define
$\psi_k\!\triangleq\!(\mathrm{x}_k, \theta_k)$.

\vspace*{0.5mm}
The transition probabilities between the operational modes 
$i,j \!\in\! \mathbb{M}$ of an MJLS are defined as

\vspace*{-5.5mm}
\begin{small}
\begin{equation}\label{eq:transition_probability_main_property}
p_{ij}\!\left(k\right) \!\triangleq\!
	\mathrm{Pr}\!\left(\theta_{k+1} \!=\! j \mid \theta_k \!=\! i  \right) 
	\!\geq\! 0, \!\!\quad \sum\nolimits_{j=1}^N p_{ij}(k) \!=\! 1.
\end{equation}
\end{small}

\vspace*{-5mm}
Since the probability distribution of a random jump variable $\theta_k$ is its 
probability mass function, $\forall i \!\in\! \mathbb{M}$ we have that
$p_i(k) \!\triangleq\!	\mathrm{Pr}\!\left( \theta_k \!=\! i \right)$ and
the initial probability distribution is then denoted
by a (column) vector
$\mathrm{p}_0 \!\triangleq\! [ p_i(0) ]_{i=1}^{N}$. 
When the initial operational mode $\theta_0$ is known 
(to be $\vartheta \!\in\!\mathbb{M}$), the information on $\mathrm{p}_0$ may 
be omitted, considering that $p_{\vartheta}(0)\!=\!1$ almost surely.
Clearly, the distribution $\mathrm{p}_k$ of 
$\theta_k$ evolves according to the transition probabilities, i.e.

\vspace*{-2mm}
\begin{small}
\begin{equation}\label{eq:distribution-evolution}
	p_j(k\!+\!1) \!=\! 
		\sum\nolimits_{i = 1}^{N} p_i(k) p_{ij}(k).
\end{equation}
\end{small}

\vspace*{-6mm}
\subsection{Polytopic time-inhomogeneous model}
In this paper we assume that the transition probabilities are \textbf{unknown} 
and \textbf{time-varying} within a bounded set. 
\begin{assumption}\label{assumption:polytope}
The transition probability matrix (TPM) $P(k)\!=\![p_{ij}(k)]_{i,j=1}^N$ 
is \textbf{polytopic time-inhomogeneous}, i.e.

\vspace*{-3.5mm}
\begin{small}
\begin{equation}\label{eq:polytope}
   P(k) \!=\! \sum\nolimits_{v=1}^V \!\lambda_v(k) P_v, \; 
   \lambda_v(k)\!\geq\! 0, \;
   \sum\nolimits_{v=1}^V \lambda_v(k) \!=\! 1,
\end{equation}
\end{small}
\vspace*{-2mm}


\vspace*{-3mm}
\noindent 
$\forall k \in \mathbb{Z}_0$,
where $V$ is a number of vertices of a convex polytope of TPMs 
$P_v\!=\![p_{ij}^{(v)}]_{i,j=1}^N$, and $\lambda_v(k)$ are unmeasurable. 
\end{assumption}

We denote by $P_{i\bullet}(k)$ the $i$-th row of 
$P(k)$, which by Assumption~\ref{assumption:polytope} belongs to a polytopic set of 
stochastic vectors. This notation allows us to denote $P(k)$ as 
$P_{[1,N]\bullet}(k)$, underlining the fact that in this case the matrix is 
interpreted row by row. We also slightly abuse our notation by indicating 
transition probability sequences of length $T$ as 
${P}_{\theta\bullet} \!\triangleq\! 
\left( \!{P}_{\theta_t \bullet}(t) \!\right)_{t=0}^{T-1}$.

\subsection{Finite-horizon optimal control}
Following the line of \cite{8264642}, it is immediate to verify that
the solution to the mode-dependent quadratic optimal control problem for PTI 
MJLSs described above, in finite-horizon case can be obtained from a 
state-dependent set of coupled difference Riccati equations (CDREs).
Specifically, when random variables 
$\left\{\mathrm{x}_t, \theta_t\right\}_{t=0}^k$ are available to the controller 
and generate a $\sigma$-algebra $\mathcal{F}$, so 
$\mathcal{F}_k \subset \mathcal{F}_{k+1} \subset \mathcal{F}$,
the optimal $\mathcal{F}_k$-measurable state-feedback controller 
$\mathrm{u}\!\triangleq\!\left( \mathrm{u}_t \right)_{t=0}^{T-1}$ that 
minimizes the quadratic functional cost associated to the closed loop system over 
a finite-time horizon, for a worst possible sequence of transition probabilities 
between the operational modes, is obtained as follows.

Let ${\bm Z} \!\triangleq\! ( Z_i )_{i=1}^{N}$, with 
$Z_i \!\in\! \mathbb{F}^{\,n_{\mathrm{x}},n_{\mathrm{x}}}_0$, be a sequence of 
the terminal cost weighting matrices. Then the optimal cost of robust control for 
the horizon of length $T$ is defined as

\vspace*{-5mm}
\begin{small}
\begin{equation}\label{eq:cost-total}
\mathcal{J}_{_{\!T\!}}(\phi) \triangleq 
	\min_{\mathrm{u}} \max_{{P}_{\theta\bullet}}\!
	\sum\nolimits_{k=0}^{T-1}\!\mathbb{E} \!\left(\|\mathrm{z}_k \|^2\right) + 
	\mathbb{E}\!\left( \mathrm{x}_T^* Z_{\theta_T} \mathrm{x}_T \right).
\end{equation}
\end{small}

\vspace*{-5mm}
Intuitively, the notation $\mathcal{J}_{_{\!T\!}}(\phi)$ 
indicates that 
the the optimal cost is attained in $T$ time steps, by starting from $\phi$.

By a standard result of dynamic programming \cite{bertsekas1995dynamic}, we have 
that a generic cost at time step $k$, when there are $T\!-\!k$ time steps left 
to the end of the control time horizon, is 

\vspace*{-5mm}
\begin{small}
\begin{equation}\label{eq:cost-to-go-generic}
J_{_{\!T-k\!}}\!\left(\psi_k,\mathrm{u}_k,{P}_{\theta_k \bullet}(k)\right) 
	\!=\!	\mathbb{E}\!\left( \!\left\| \mathrm{z}_k \right\|^{2} \!+\! 
	\mathcal{J}_{_{\!T-k-1\!}}\!\left(\psi_{k+1}\right) \!\mid\! 
	\mathcal{F}_k \!\right),
\end{equation}
\end{small}
\vspace*{-6.5mm}

\noindent where the cost-to-go function is defined as

\vspace*{-3mm}
\begin{small}
\begin{equation}\label{eq:cost-to-go-optimal}
\mathcal{J}_{_{\!T-k\!}}\!\left(\psi_k \right) \!=\! 
\min_{\mathrm{u}_k} \max_{{P}_{\theta_k \bullet}(k)} 
J_{_{\!T-k\!}}\!\left(\psi_k,\mathrm{u}_k,\!{P}_{\theta_k\bullet}(k)\right).
\end{equation}
\end{small}

\vspace*{-5mm}
By the definition of the expected value, we have 
from \eqref{eq:cost-to-go-generic} that
the cost-to-go function \eqref{eq:cost-to-go-optimal} is equal to

\vspace*{-4.5mm}
\begin{small}
\begin{equation}\label{eq:cost-to-go-optimal_expected}
\mathcal{J}_{_{\!T-k\!}}\!\left(\psi_k \right) \!=\! 
	\min_{\mathrm{u}_k} \!\!\max_{P_{\left[1,N\right] \bullet}\!(k)}\!
	\sum\nolimits_{i=1}^{N} \!p_i(k) J_{_{\!T-k\!}}\!
	\left(\psi_k, \!\mathrm{u}_k,\!P_{\!i \bullet}(k) \!\right),
\end{equation}
\end{small}

\vspace*{-4.5mm}
\noindent
where we emphasize the fact that the cost-to-go function is determined by a set 
of $N$ generic costs, each one associated to a different row of the same TPM $P(k)$.

We assume without loss of generality 
\cite[p.74, Remark 4.1]{costa2006discrete} 
that for all $i \!\in\! \mathbb{M}$

\vspace*{-9.5mm}
\begin{small}
\begin{equation}\label{assumption:c_i_d_i}
\qquad\qquad\qquad\qquad\quad C_i^* D_i \!=\! 0, \quad D_i^* D_i \!\in\!\mathbb{F}^{n_{\mathrm{u}},n_{\mathrm{u}}}_0.
\end{equation}
\end{small}

\vspace*{-6mm}
\noindent 
Then via some manipulations described in \cite{8264642} and 
explained in detail in \cite{yzl_dissertation}, we 
can prove that at each time step $k$, the maximum in transition probabilities 
of a generic cost \eqref{eq:cost-to-go-generic}
is attained on one of the $V$ vertices of the convex polytope of stochastic matrices $P_{v}$
that bound the values of the uncertain and time-varying TPM $P(k)$. 
Thus, to find a worst possible sequence of transition probabilities we need 
to consider only the TPMs corresponding to the vertices $P_{v}$. For each of these vertices,
the minimum in $\mathrm{u}_k$ of a generic cost \eqref{eq:cost-to-go-generic} is achieved
by a state-feedback controller derived from the solution to CRDEs that is obtained by the following backward recursion:

\vspace*{-4.5mm}
\begin{small}
\begin{equation}\label{eq:r_i_k}
R_{i}^{(v_l)}\!(k) = \left(\!D_{i}^* D_{i} \!+\! 
	B_{i}^* \sum\nolimits_{j=1}^N p_{ij}^{(v)} X_j^{(l)}\!(k\!+\!1) B_{i}\! \right)^{\!-1} ,
\end{equation}
\vspace*{-2mm}
\begin{equation}\label{eq:k_i_k}
K_{i}^{(v_l)}\!(k) = - R_{i}^{(v_l)}\!(k)
B_{i}^* \!\sum\nolimits_{j=1}^N p_{ij}^{(v)} X_j^{(l)}\!(k\!+\!1) A_{i} ,
\end{equation}
\vspace*{-4mm}
\begin{align}\label{eq:X_i_k}
X_{i}^{(v_l)}\!(k) \triangleq 
& \, C_i^* C_i + A_i^* \! \sum\nolimits_{j=1}^N p_{ij}^{(v)} X_j^{(l)}\!(k\!+\!1) A_i + \nonumber \\[-1mm]
& \, A_i^* \!\sum\nolimits_{j=1}^N p_{ij}^{(v)} X_j^{(l)}\!(k\!+\!1) B_i K_{i}^{(v_l)}\!(k) ,
\end{align}
\end{small}

\vspace*{-4mm}
\noindent where the superscript $(v_l)$ indicates an element obtained with 
transition probabilities corresponding to vertex $v\!\leq\!V$ 
from the $l$-th solution to CDRE at the next time step, $l\!\leq\!L_{k+1}$.
At the last time step we have $L_T\!=\!1$, so that $X_i^{(1)}\!(T)\!=\!Z_i$.

To each ${\bm X}_{v_l}(k)\!\triangleq\!(X_i^{(v_l)}\!(k))_{i=1}^{N}$ corresponds a cost

\vspace*{-1.5mm}
\begin{small}
\begin{equation}\label{eq:cost-to-go_total_explicit_k}
\mathcal{J}^{(v_l)}_{_{\!T-k\!}}\!\left(\psi_{k}\right) =
	\mathrm{x}_{k}^* \!\left( \sum\nolimits_{i=1}^{N} \! p_i^{(v)}\!(k) X_{i}^{(v_l)}\!(k)\!\right)\! \mathrm{x}_{k} ,
\end{equation}
\end{small}

\vspace*{-5mm}
\noindent
where $p_i^{(v)}\!(k)$ is obtained by forward recursion starting from the initial probability distribution 
$\mathrm{p}_0$ via \eqref{eq:distribution-evolution}, i.e.,

\vspace*{-2.5mm}
\begin{small}
\begin{equation}\label{eq:distribution-evolution-vertex}
p_i^{(v)}(k) \!=\! \sum\nolimits_{j = 1}^{N} p_j(k\!-\!1) p_{ji}^{(v)} .
\end{equation}
\end{small}

\vspace*{-5mm}
Then, for any $\mathrm{x}_k$, the cost-to-go is obtained simply as

\vspace*{-3mm}
\begin{small}
\begin{equation}\label{eq:cost-to-go-optimal_expected_k}
\mathcal{J}_{_{\!T-k\!}}\!\left(\psi_k \right) = \max_{v_l} \mathcal{J}^{(v_l)}_{_{\!T-k\!}}\!\left(\psi_{k}\right) .
\end{equation}
\end{small}

\vspace*{-5mm}
After denoting by $\bar{v}_k = \arg \max\limits_{v_l} \mathcal{J}_{_{\!T-k\!}}\!\left(\psi_k \right)$ 
the index of the solution to CDRE corresponding to the cost-to-go, we can write the optimal control input as

\vspace*{-2mm}
\begin{small}
\begin{equation}\label{eq:u_k}
\mathrm{u}_{k} = K_{\theta_k}^{(\bar{v}_k)}\!(k) \, \mathrm{x}_{k} .
\end{equation}
\end{small}

\vspace*{-6mm}
Since the terminal cost weighting matrices and the 
set of vertices of the convex polytope that bounds the values of TPM are known, the 
solutions to CRDEs \eqref{eq:r_i_k}\,--\,\eqref{eq:X_i_k} can be computed off-line.
The online controller relies on \eqref{eq:cost-to-go_total_explicit_k}\,--\,\eqref{eq:u_k} and the 
number of solutions to CRDEs to consider may act as a bottleneck, 
since at $T\!-\!k$ time steps left to the end of the control time horizon, there are $V^{k}$
options to consider.

Nevertheless, some costs \eqref{eq:cost-to-go-optimal_expected_k}
obtained from solutions ${\bm X}_{v_l}(k)\!\triangleq\!(X_i^{(v_l)}\!(k))_{i=1}^{N}$,
$X_i^{(v_l)}\!(k)\!\in\!\mathbb{F}^{n_{\mathrm{x}},n_{\mathrm{x}}}_0$ may not achieve
maximum in transition probabilities and thus will never lead to cost-to-go \eqref{eq:cost-to-go-optimal}.
Such solutions are redundant and could be discarded. So, the set of all $V L_{k+1}$ solutions
${\bm X}_{v_l}(k)$ can be pruned, obtaining a so-called parsimonious set of $L_k \!\leq\!V L_{k+1}$ elements.
Specifically, the set $\{{\bm X}_l(k)\}_{l=1}^{L_k}$ of solutions to CDREs is said to be parsimonious
if there is \textbf{no other} solution ${\bm X}_{\ell}(k)$ such that
\vspace*{-1.5mm}
\begin{small}
\begin{equation}\label{eq:parsimonious}
X_i^{(\ell)}\!(k)\!-\!X_i^{(l)}\!(k) \in\mathbb{F}^{n_{\mathrm{x}},n_{\mathrm{x}}}_0 \quad
\forall i \!\in\!\mathbb{M},\, l\!\leq\!L_k .
\end{equation}
\end{small}

\vspace*{-5mm}
Note that, if satisfied, \eqref{eq:parsimonious} means that 
\begin{small}
\begin{equation*}
	 \sum\nolimits_{i=1}^{N} \! p_i(k) 
	 \mathrm{x}_{k}^* \left( X_{i}^{(\ell)}\!(k) - X_{i}^{(l)}\!(k) \right) \mathrm{x}_{k} \!\geq\! 0 
	 \quad \forall \mathrm{x}_{k}, \mathrm{p}_k ,
\end{equation*}
\end{small}

\vspace*{-4.2mm}
\noindent
so the solution ${\bm X}_{l}(k)$ would be redundant, to be discarded. 
We observe that conditions for pruning of redundant solutions, that are similar to \eqref{eq:parsimonious},
were already established for switched Riccati mapping associated with quadratic regulation problem 
for discrete-time switched linear systems \cite{zhang2012infinite}.
The number of remaining parsimonious solutions 
depends on the rate of convergence of the closed-loop system, as will be discussed in the next sections.

\subsection{Infinite-horizon optimal control problem}
When the control horizon $T$ goes to infinity, the terminal cost will never be incurred, so
the optimal cost of robust control \eqref{eq:cost-total} becomes 

\vspace*{-2.5mm}
\begin{small}
\begin{equation}\label{eq:cost-total-infty}
\mathcal{J}_{_{\!\infty\!}}(\phi) \triangleq \min_{\mathrm{u}} \max_{{P}_{\theta\bullet}} 
	\!\sum\nolimits_{k=0}^{\infty} \!\mathbb{E} \!\left( \| \mathrm{z}_k \|^2 \right) .
\end{equation}
\end{small}

\vspace*{-4.5mm}
\noindent
In this paper, we are interested in the steady-state stabilizing solution to the optimal state-feedback control problem,
i.e., to find analytical expression of $\mathrm{u}\!=\!\left( \mathrm{u}_k \right)_{k=0}^{\infty}$ that achieves 
\eqref{eq:cost-total-infty} and stabilizes the MJLS \eqref{eq:model_mjls_control} in the mean square sense.
\section{MEAN SQUARE STABILIZABILITY}\label{sec:mss}
\noindent
We recall from \cite[pp. 36\,--\,37, Definition 3.8, Remark 3.10]{costa2006discrete} that 
an autonomous MJLS 
$\mathrm{x}_{k+1}\!=\!A_{\theta_k} \mathrm{x}_k$ is mean square stable (MSS)
if for any initial conditions $\mathrm{x}_0$ and $\theta_0$, one has 

\vspace*{-2.5mm}
\begin{small}
\begin{equation} \label{eq:mss_noiseless}
   \lim_{k \to \infty} \mathbb{E}\!\left( \mathrm{x}_k \right) = 0, \qquad
   \lim_{k \to \infty} \mathbb{E}\!\left( \mathrm{x}_k \mathrm{x}_k^* \right) = 0 .
\end{equation}
\end{small}

\vspace*{-5mm}
In \cite[Theorems 1 and 3]{yzl2016cdc} we have proved that in PTI setting provided by 
Assumption~\ref{assumption:polytope}, the system 
is MSS if and only if the joint spectral radius (see \cite{jungers2009joint} 
and references therein for an overview) of a finite family of matrices (associated to the 
second moment of the MJLS) is smaller than 1, and that the mean square stability is equivalent 
to the exponential mean square stability (EMSS) and to the stochastic stability (SS). 
Formally, let $_{\mathbb{V}}{\bm \Lambda}\!\triangleq\!\{\Lambda_v\}_{v=1}^{V}$ 
be a set of matrices related to the second moment of $\mathrm{x}_k$, with 

\vspace*{-2.5mm}
\begin{small}
\begin{equation}\label{eq:second_moment_matrix}
\Lambda_v \triangleq \left( P_v^T \otimes I_{n_{\mathrm{x}}^2}\right) 
\left( \bigoplus\nolimits_{i=1}^{N} \left( \bar{A}_i \otimes A_i \right) \right) .
\end{equation}
\end{small}

\vspace*{-5mm}
\noindent
Then, for a PTI MJLS as in \eqref{eq:model_mjls_control}, \eqref{eq:polytope}, the following statements are \textbf{equivalent}:
\begin{enumerate}
\item $\hat{\rho}(_{\mathbb{V}}{\bm \Lambda}) \!<\! 1$;
\item the system is MSS, i.e., it satisfies \eqref{eq:mss_noiseless} $\forall \mathrm{x}_0,\theta_0$;
\item the system is EMSS, i.e., $\forall k\!\in\!\mathbb{Z}_0,\mathrm{x}_0,\theta_0$, and 
for some reals $\beta \!\geq\! 1$, $0 \!<\! \zeta \!<\! 1$, one has that 

\vspace*{-2.5mm}
\begin{small}
\begin{equation}\label{eq:emss}
\mathbb{E}\!\left( \left\| \mathrm{x}_k \right\|^2 \right) \leq \beta \zeta^k \left\| \mathrm{x}_0 \right\|^2 ;
\end{equation}
\end{small}

\vspace*{-6mm}
\item the system is SS, i.e.~$\forall \mathrm{x}_0,\theta_0$, one has that

\vspace*{-2.5mm}
\begin{small}
\begin{equation}\label{eq:ss}
\sum\nolimits_{k=0}^{\infty}
\mathbb{E}\!\left( \left\| \mathrm{x}_k \right\|^2 \right) < \infty .
\end{equation}
\end{small}
\end{enumerate}

As in \cite[p. 57, Definition 3.40]{costa2006discrete}, we will say that the pair 
$({\bm A}, {\bm B})$ of $N$-sequences of state and control input matrices 
related to all operational modes of the system \eqref{eq:model_mjls_control},
is \textbf{mean square stabilizable} if there exists a sequences of control matrices 
${\bm K} \!=\! (K_i)_{i=1}^{N}$ 
such that the system \eqref{eq:model_mjls_control} with synchronous state feedback controller 
$\mathrm{u}_k = K_{\theta_k} \mathrm{x}_k$ is MSS. In this case, ${\bm K}$ is said to stabilize 
the pair $({\bm A},{\bm B})$.

Since the controller $\mathrm{u}_k \!=\! K_{\theta_k} \mathrm{x}_k$ 
gives to the system \eqref{eq:model_mjls_control}
an autonomous form, i.e.
$\mathrm{x}_{k+1}
\!=\! \left( A_{\theta_k} \!+\! B_{\theta_k} K_{\theta_k} \right) \mathrm{x}_k 
\!=\!\mathit{\Gamma}_{\!\theta_k}\mathrm{x}_k$,
we can apply the previous result on stability equivalence to the controlled
system, after substituting in \eqref{eq:second_moment_matrix} 
$A_i$ with $\mathit{\Gamma}_i\!\triangleq\!( A_{i} \!+\! B_{i} K_{i})$, $\forall i \!\in\!\mathbb{M}$.
Specifically, 
let $_{\mathbb{V}}{\bm \Delta}\!\triangleq\!\{\Delta_v\}_{v=1}^{V}$,
$\Delta_v \!\triangleq\! \left( P_v^T \!\otimes\! I_{n_{\mathrm{x}}^2}\right)\!\! 
	\left( \bigoplus\nolimits_{i=1}^{N} \left(\,\bar{\!\mathit{\Gamma}}_i \!\otimes\! \mathit{\Gamma}_i \right) \right)$,
be a set of matrices related to the second moment of the controlled state 
$\mathrm{x}_k$, then the system \eqref{eq:model_mjls_control} is stabilizable 
if and only if there exists a sequence ${\bm K}\!=\!(K_i)_{i=1}^{N}$ 
of control matrices such that $\hat{\rho}(_{\mathbb{V}}{\bm \Delta}) \!<\! 1$.

We observe that, while it is NP-hard to decide whether 
$\hat{\rho}(_{\mathbb{V}}{\bm \Delta}) \!<\! 1$ \cite[Theorem 2]{yzl2016cdc},
there exist computationally efficient approximations of the joint spectral radius (\cite{jungers2009joint,blondel2005computationally,protasov2010joint,vankeerberghen2014jsr}),
allowing us to verify whether the system is stabilized by a given $N$-sequence ${\bm K}$ of
control matrices.

For the sake of completeness, we extend the definition of the
mean square detectability \cite[p. 57, Definition 3.41]{costa2006discrete} to 
PTI MJLSs by stating that the pair $({\bm C},{\bm A})$ of $N$-sequences of 
output and state matrices related to all operational modes of the system 
\eqref{eq:model_mjls_control} is \textbf{mean square detectable} if there 
exists a sequence of filter gain matrices 
${\bm G} \!\triangleq\!\left( G_i \right)_{i=1}^{N}$, 
$G_i \!\in\!\mathbb{F}^{n_{\mathrm{x}},n_{\mathrm{z}}}$, such that
for ${\bm F} \!\triangleq\!\left( F_i \right)_{i=1}^{N}$,
$\mathit{F}_i\!\triangleq\!( A_{i} \!+\! G_{i} C_{i} )$,
$_{\mathbb{V}}{\bm \Phi}\!\triangleq\!\{\Phi_v\}_{v=1}^{V}$,
$\Phi_v \!\triangleq\! \left( P_v^T \!\otimes\! I_{n_{\mathrm{x}}^2}\right) 
\!\!\left( \bigoplus\nolimits_{i=1}^{N}\!
\left(\,\bar{\!\mathit{F}}_i \otimes \mathit{F}_i \right) \right)$,
one has that $\hat{\rho}(_{\mathbb{V}}{\bm \Phi}) \!<\! 1$.
This definition ensures that the evolution of the observation error for the 
system \eqref{eq:model_mjls_control} with null direct transition matrices and
the synchronous full-order Markov jump filter having a structure similar to the 
structure of the Luenberger observer is mean square stable \cite{yzl_dissertation}.
We will see in the next section that the mean square detectability
(defined explicitly for PTI MJLSs) ensures that the stabilizing solution to a set 
of coupled algebraic Riccati equations exists and is unique.

As a final remark, we underline that the mean square 
stabilizability and detectability tests 
\cite[pp. 57--59, Propositions 3.42, 3.43]{costa2006discrete}
for time-homogeneous MJLSs can be easily extended to PTI case, with the 
requirement that the linear matrix inequalities (LMIs) are satisfied 
for all vertices of the polytopic TPM. Obviously, these tests provide necessary,
but not sufficient conditions, so the feasible solutions ${\bm K}$ and ${\bm G}$
should be tested also through $\hat{\rho}(_{\mathbb{V}}{\bm \Delta})$ and
$\hat{\rho}(_{\mathbb{V}}{\bm \Phi})$.
\section{COUPLED ALGEBRAIC RICCATI EQUATIONS}\label{sec:care}
When the MJLS is governed by a Markov chain with a stationary TPM, which is 
known exactly, if the system is 
mean square stabilizable and detectable, then 
there exists the mean square stabilizing solution for the set of coupled 
algebraic Riccati equations (CAREs) that provides an optimal state-feedback 
control law and achieves the optimal cost. Furthermore, in this case the solution 
of CDREs converges to the unique solution of the related CAREs, which coincides 
with the maximal solution of the same problem and it can be obtained numerically 
via a certain LMI optimization problem 
\cite[pp. 78\,--\,81, 203\,--\,228]{costa2006discrete}.

In the rest of this paper we generalize the aforementioned result to MJLSs with 
polytopic time-inhomogeneous 
transition probabilities, as by Assumption~\ref{assumption:polytope}.

\begin{definition}\label{definition:control_care}
We say that the
$\{\hat{{\bm X}}_l\}_{l=1}^{L}$, 
$\hat{{\bm X}}_l\!\triangleq(\hat{X}_i^{(l)})_{i=1}^{N}$,
$\hat{X}_i^{(l)}\!\in\!\mathbb{F}^{n_{\mathrm{x}},n_{\mathrm{x}}}_0$, $L\!\leq\!V$,
is the \textbf{stabilizing solution} for the \textbf{control CAREs} 
associated to MJLS \eqref{eq:model_mjls_control} in PTI setting \eqref{eq:polytope} if 
$\forall i\!\in\!\mathbb{M}$ and $\forall l\!\leq\!L$,
it satisfies

\vspace*{-2.5mm}
\begin{small}
\begin{equation}\label{eq:r_i}
\hat{R}_{i}^{(l)} = \left(\!D_{i}^* D_{i} \!+\! 
	B_{i}^* \sum\nolimits_{j=1}^N p_{ij}^{(l)} \hat{X}_j^{(l)} B_{i}\! \right)^{\!-1} ,
\end{equation}

\vspace*{-2.5mm}
\begin{equation}\label{eq:k_i}
\hat{K}_{i}^{(l)} = - \hat{R}_{i}^{(l)} B_{i}^* \!\sum\nolimits_{j=1}^N p_{ij}^{(l)} \hat{X}_j^{(l)} A_{i} ,
\end{equation}
\begin{equation}\label{eq:X_i}
\hat{X}_{i}^{(l)} \!\!=\!
C_i^* C_i \!+\! A_i^* \!\! \sum\nolimits_{j=1}^N p_{ij}^{(l)} \!\hat{X}_j^{(l)}\!\! A_i \!+\!
A_i^* \!\!\sum\nolimits_{j=1}^N \!p_{ij}^{(l)}\! \hat{X}_j^{(l)}\! B_i \hat{K}_{i}^{(l)}\!\!,
\end{equation}
\end{small}

\vspace*{-4mm}
\noindent
and $\forall\mathit{\Gamma}_i^{(l)}\!\triangleq\!\left( A_{i} \!+\! B_{i} \hat{K}_{i}^{(l)} \right)$,
$_{\mathbb{V}}{\bm \Delta}^{\!(l)}\!\triangleq\!\left\{\Delta_{v}^{\!(l)}\right\}_{v=1}^{V}$,
with

\vspace*{-0.5mm}
\begin{small}
\begin{equation}\label{eq:second_moment_l}
\Delta_v^{(l)} \triangleq \left( P_v^T \otimes I_{n_{\mathrm{x}}^2}\right) 
	\left( \bigoplus\nolimits_{i=1}^{N} \left(\,\bar{\!\mathit{\Gamma}}_i^{(l)} \otimes \mathit{\Gamma}_i^{(l)} \right) \right) ,
\end{equation}
\end{small}

\vspace*{-4mm}
\noindent
it satisfies $\hat{\rho}(_{\mathbb{V}}{\bm \Delta}^{\!(l)}) \!<\! 1$, and if $L\!<\!V$, every
solution $\hat{\bm X}_{\ell}$, $L\!<\!\ell\!\leq\!V$ satisfying \eqref{eq:r_i}\,--\,\eqref{eq:X_i} is such that for some $l\!\leq\!L$

\vspace*{-2mm}
\begin{small}
\begin{equation}\label{eq:parsimonious_inf}
\hat{X}_i^{(l)}\!-\!\hat{X}_i^{(\ell)} \in\mathbb{F}^{n_{\mathrm{x}},n_{\mathrm{x}}}_0 \quad
\forall i \!\in\!\mathbb{M} .
\end{equation}
\end{small}
\end{definition}

\vspace*{1.5mm}
In words, Definition \ref{definition:control_care} states that, when the 
TPM
of an MJLS is unknown and time-varying within a bounded set as by Assumption~\ref{assumption:polytope},
instead of one set of $N$ coupled algebraic Riccati equations, there are up to $V$ sets of CAREs, each 
one associated to a different vertex of a convex polytope of transition probabilities. If a solution
corresponding to some vertex $\ell$ satisfies \eqref{eq:parsimonious_inf}, such solution is redundant 
and should not be considered, because it will never give the optimal cost \eqref{eq:cost-total-infty}.
All the remaining solutions form a parsimonious set, since they produce the smallest set of solutions of CAREs
that achieves the optimal cost \eqref{eq:cost-total-infty} for any value of the initial state $\mathrm{x}_0$.
So the state space is partitioned in $L\!\leq\!V$ regions, each one with a different optimal control law.
Each controller will stabilize the system if and only if the joint spectral radius of
a set of matrices related to the second moment of the controlled system's state is smaller than 1, as reiterated
in the central part of the definition. When existing, the defined stabilizing solution for the control CAREs is \textbf{unique}, because
from \eqref{eq:r_i}\,--\,\eqref{eq:X_i} each $\hat{{\bm X}}_l$ associated to a partition of state-space is a 
parsimonious solution to a set of CAREs where the transition probabilities are stationary and known, 
so each $\hat{{\bm X}}_l$ is unique and can be computed numerically 
\cite[p. 215]{costa2006discrete}:

\vspace*{-2mm}
\begin{small}
\begin{equation*}
\hat{{\bm X}}_l = \arg \max_{{\bm X}_l} \mathrm{tr} \left(\sum\nolimits_{i=1}^{N}\!X_i^{(l)}\right)~\text{subject to}
\end{equation*}
\vspace*{-4mm}
\begin{equation*}
{\bm X}_l \!=\! \left(\!X_{i}^{(l)}\!\right)_{i=1}^{N}\!, \text{ where } 
\forall i\!\in\!\mathbb{M},\quad  X_{i}^{(l)} \!=\! \left(\!X_{i}^{(l)}\!\right)^{\!T}, \text{ and}
\end{equation*}
\vspace*{-4mm}
\begin{equation*}
D_{i}^* D_{i} \!+\!\! B_{i}^*\!\! \sum\nolimits_{j=1}^N \!p_{ij}^{(l)} \! X_j^{(l)}\! B_{i}\!\succ\!0,
\end{equation*}
\vspace*{-2.5mm}
\begin{equation*}\label{eq:max_care_lmi}
\begin{bmatrix}
\!-\!X_i \!+\!\! A_i^* \! \sum\nolimits_{j=1}^N \!p_{ij}^{(l)} \!X_j^{(l)}\!\! A_i\!+\! C_i^* C_i & 
A_i^* \! \sum\nolimits_{j=1}^N \!p_{ij}^{(l)} \!X_j^{(l)}\! B_i \\[2mm]
\!\!\!\!\!\!\!\!\!\!\!\!\!\!
B_i^* \! \sum\nolimits_{j=1}^N \!p_{ij}^{(l)} \!X_j^{(l)}\!\! A_i & 
\!\!\!\!\!\!\!\!\!
D_{i}^* D_{i} \!+\!\! B_{i}^*\! \sum\nolimits_{j=1}^N \!p_{ij}^{(l)} \! X_j^{(l)}\! B_{i}\!
\end{bmatrix} \!\!\succeq\!0.
\end{equation*}
\end{small}

\vspace*{-2mm}
\noindent The LMI optimization problem above can be easily implemented in Matlab-based Robust Control Toolbox
\cite{balas2017robust} and solved via its 
solver \texttt{mincx}, 
as illustrated in Section \ref{sec:example}.

It is worth mentioning that if the solution to the above LMI optimization problem 
exists for an $l\!\leq\!V$, then 
the spectral radius
of the matrix
$\Delta_{l}^{\!(l)}$ associated to the second moment of the MJLS with TPM $P_l$ is
less then one,
which is a necessary but not 
sufficient condition for having 
$\hat{\rho}(_{\mathbb{V}}{\bm \Delta}^{\!(l)}) \!<\! 1$. Thus, the last 
condition should be checked separately 
$\forall \hat{\bm K}_l\!=\! (\!\hat{K}_i^{(l)})_{i=1}^{N}$. 

We stress that 
the mean square detectability of the pair $({\bm C},{\bm A})$
for a PTI MJLS implies the mean square detectability
for a MJLS with stationary TPM $P_{l}$ (since $\rho({\Phi_v})\!\leq\!\hat{\rho}(_{\mathbb{V}}{\bm \Phi})$), 
which in turn ensures that the stabilizing solution 
to a set of CAREs exists if the system is also mean square stabilizable 
\cite[p. 218, Corollary A.16]{costa2006discrete}.
By \cite[p. 42, Proposition 3.20]{costa2006discrete} this solution is unique.

Let us define the robust control cost function associated to the stabilizing solution for the control CAREs as 

\vspace*{-2.5mm}
\begin{small}
\begin{equation}\label{eq:cost-to-go-infty}
\hat{\mathcal{J}}_{_{\!\infty\!}}(\psi_k) \!=\! \max\limits_l
\mathrm{x}_{k}^* \!\left( \sum\nolimits_{i=1}^{N} \! p_i^{(l)}\!(k) \hat{X}_{i}^{(l)}\!\right)\! \mathrm{x}_{k},
\end{equation}
\end{small}

\vspace*{-3mm}
\noindent
with
$p_i^{(l)}\!(k)$ computed via \eqref{eq:distribution-evolution-vertex}, and

\vspace*{-3mm}
\begin{small}
\begin{equation}\label{eq:hat_v_k}
\hat{v}_k\!=\!\arg \max\limits_{l} \hat{\mathcal{J}}_{_{\!\infty\!}}(\psi_k) .
\end{equation}
\end{small}

\vspace*{-5mm}
\noindent
Then we can present the first main result of this paper. 

\begin{theorem}\label{theorem:care}
Suppose that the stabilizing solution 
$\{\!\hat{{\bm X}}_l\}_{\!l=1}^{\!L}\!\!$ for the control CAREs 
associated to MJLS \eqref{eq:model_mjls_control} in PTI setting 
\eqref{eq:polytope} exists. Then the control law 
$\hat{\mathrm{u}}\!=\!\left( \hat{\mathrm{u}}_k \right)_{k=0}^{\infty}$, where 
$\hat{\mathrm{u}}_k\!=\!\hat{K}_{\theta_k}^{(\hat{v}_k)}\mathrm{x}_k$, 
stabilizes the system in the mean square sense and achieves the optimal cost 
\eqref{eq:cost-total-infty}, which is given by 
$\hat{\mathcal{J}}_{_{\!\infty\!}}(\phi)$.

\begin{proof}
Since the solution 
$\{\!\hat{{\bm X}}_l\}_{l=1}^{L}\!$ of the control CAREs is 
stabilizing, by Definition \ref{definition:control_care} it gives us a 
set $\{\!\hat{{\bm K}}_l\}_{l=1}^{L}$, where
$\hat{\bm K}_l\!=\! (\!\hat{K}_i^{(l)})_{i=1}^{N}$ and $\hat{K}_i^{(l)}$ is
provided by \eqref{eq:k_i}, such that 
$\hat{\rho}(_{\mathbb{V}}{\bm \Delta}^{\!(l)}) \!<\! 1$ $\forall l$.
So, the system \eqref{eq:model_mjls_control} with TPM as in \eqref{eq:polytope} 
and the control law 
$\hat{\mathrm{u}}\!=\!\left( \hat{\mathrm{u}}_k \right)_{k=0}^{\infty}$,
$\hat{\mathrm{u}}_k\!=\!\hat{K}_{\theta_k}^{(\hat{v}_k)}\mathrm{x}_k$,
is MSS. Thus, by the stability equivalence, the system is also stochastically 
stable and
$\lim\nolimits_{k\to\infty}\mathbb{E}\!\left(\|\mathrm{x}_k\|^{2}\right)\!=\!0$,
as proved in \cite{yzl2016cdc}. 
Now, for what concerns the optimal cost $\mathcal{J}_{_{\!\infty\!}}(\phi)$,
let $\mathcal{U}$ be a set of all mean square stabilizing control laws. 
For any law $\mathrm{u} \!\in\! \mathcal{U}$, $\mathrm{u} \!=\!(\mathrm{u}_{k})_{k=0}^{\infty}$, 
we have from \eqref{eq:model_mjls_control}, \eqref{eq:r_i}\,--\,\eqref{eq:X_i}, and the fact that 
the maximum in transition probabilities of the generic cost at time $k$ is attained on a 
vertex, denoted by $\hat{v}_k$, of the convex polytope of TPMs, for all $\mathrm{x}_k$ and 
$\hat{X}_{\theta_{k}}^{(l)}$, with $l\!\leq\!L\!\leq\!V$, that
\begin{small}
\begin{align*}
\max\limits_{P_{\theta_k\bullet}(k)} 
\mathbb{E}\!\left(\!
\mathrm{x}_{k+1}^{*} \hat{X}_{\theta_{k+1}}^{(l)}\!\mathrm{x}_{k+1} \!-\!
\mathrm{x}_{k}^{*} \hat{X}_{\theta_{k}}^{(l)}\mathrm{x}_{k} \!+\!
\mathrm{u}_{k}^{*} D_{\theta_{k}}^{*} \!D_{\theta_{k}} \!\mathrm{u}_{k} \!\mid\! \mathcal{F}_{k}
\!\right) \!=\! \\
\mathbb{E}\!\left(\!\!
\left(\!\mathrm{u}_{k} \!-\! \hat{K}_{\theta_{k}}^{(\hat{v}_k)}\mathrm{x}_{k}\!\right)^{\!\!*}\!\!
\left(\!\hat{R}_{\theta_{k}}^{(\hat{v}_k)}\!\right)^{\!\!-1}\!\!
\left(\!\mathrm{u}_{k} \!-\! \hat{K}_{\theta_{k}}^{(\hat{v}_k)}\mathrm{x}_{k}\!\right) \!-\!
\mathrm{x}_{k}^{*} C_{\theta_{k}}^{*} \!C_{\theta_{k}} \!\mathrm{x}_{k}\!
\!\right),
\end{align*}
\end{small}

\vspace*{-3mm}
\noindent
which, together with \eqref{eq:model_mjls_control} and \eqref{assumption:c_i_d_i}, implies that
\begin{small}
\begin{align*}
&~~~
\mathbb{E}\!\left(\|\mathrm{z}_k\|^{2}\right)\!=\!
\mathbb{E}\!\left(\mathrm{z}_k^{*}\mathrm{z}_k\right)\!=\!
\mathbb{E}\!\left(
\mathrm{x}_{k}^{*} C_{\theta_{k}}^{*} \!C_{\theta_{k}} \!\mathrm{x}_{k} \!+\!
\mathrm{u}_{k}^{*} D_{\theta_{k}}^{*} \!D_{\theta_{k}} \!\mathrm{u}_{k}
\right)\!=\!\\
&\!
\max\limits_{P_{\theta_k\bullet}(k)} \!\!\!
\mathbb{E}\!\!\left(\!\!
\mathrm{x}_{\;\!\!k}^{*} \hat{X}_{\;\!\!\theta_{\!k}}^{\!(l)}\!\mathrm{x}_{\;\!\!k} \!\!-\!\!
\mathrm{x}_{\;\!\!k+\;\!\!1}^{*} \!\hat{X}_{\;\!\!\theta_{\!k+\;\!\!1}}^{\!(l)}\!\!\!\mathrm{x}_{\;\!\!k+\;\!\!1} \!+\!
\left\|\!
\left(\!\hat{R}_{\;\!\!\theta_{\!k}}^{\;\!\!(\!\hat{v}_k\!)}\!\right)^{\!\!-\!\frac{1}{2}}\!\!\!
\left(\!\mathrm{u}_{\;\!\!k} \!\!-\!\! \hat{K}_{\;\!\!\theta_{\!k}}^{\;\!\!(\!\hat{v}_k\!)}\!\mathrm{x}_{\;\!\!k}\!\right)\!
\right\|^{2}
\right) .
\end{align*}
\end{small}

\vspace*{-3mm}
\noindent
Since for all $\mathrm{u} \!\in\!\mathcal{U}$ we have that
$\lim\nolimits_{k\to\infty}\mathbb{E}\!\left(\|\mathrm{x}_k\|^{2}\right)\!=\!0$, and, consequently,
$\lim\nolimits_{k\to\infty}\mathbb{E}(\mathrm{x}_{k}^{*} \hat{X}_{\theta_{k}}^{(l)}\mathrm{x}_{k})\!=\!0$, it follows that
\begin{small}
\begin{align*}
\max_{{P}_{\theta\bullet}} \!\sum\nolimits_{k=0}^{\infty} \!
\mathbb{E} \!\left(  
\left\|\!
\left(\!\hat{R}_{\theta_{k}}^{(\!\hat{v}_k\!)}\!\right)^{\!-\!\frac{1}{2}}\!\!
\left(\!\mathrm{u}_{k} \!-\! \hat{K}_{\theta_{k}}^{(\!\hat{v}_k\!)}\mathrm{x}_{k}\!\right)\!
\right\|^{2}
\right)\!+\!
\mathbb{E} \!\left( \!
\mathrm{x}_{0}^{*} \hat{X}_{\theta_{k}}^{(l)}\mathrm{x}_{0} \!
\right)
\end{align*}
\end{small}

\vspace*{-3mm}
\noindent
is minimized for $\mathrm{u}\!=\!\hat{\mathrm{u}}$. Then, considering also the definition of the expected value and 
\eqref{eq:model_mjls_control}, it follows that
$\mathcal{J}_{_{\!\infty\!}}(\phi)$ equals to
$\mathbb{E} \!\left(\!\mathrm{x}_{0}^{*} \hat{X}_{\theta_{0}}^{(\!\hat{v}_0\!)}\mathrm{x}_{0} \!\right)\!=\!
\mathrm{x}_{0}^* \!\left( \sum\nolimits_{i=1}^{N} \! p_i(0) \hat{X}_{i}^{(\!\hat{v}_0\!)}\!\right)\! \mathrm{x}_{0} \!=\!
\hat{\mathcal{J}}_{_{\!\infty\!}}(\phi)$.
\end{proof}
\end{theorem}

\vspace*{1mm}
We observe that, by stability equivalence, the MJLS with control law 
$\left( \hat{\mathrm{u}}_k \right)_{k=0}^{\infty}$, 
$\hat{\mathrm{u}}_k\!=\!\hat{K}_{\theta_k}^{(\hat{v}_k)}\mathrm{x}_k$, 
is EMSS, so it generates a trajectory satisfying \eqref{eq:emss} 
$\forall k\!\in\!\mathbb{Z}_0$, with e.g.~

\vspace*{-2mm}
\begin{small}
\begin{equation}\label{eq:emss_bound_on_x}
\max_{l}\hat{\rho}(_{\mathbb{V}}{\bm \Delta}^{\!(l)})\!\leq\!\zeta\!<\! 1 
\text{~and~} \beta\!\geq\!n_\mathrm{x} N, ~\forall k\!\geq\!k'\!\geq\!0 .
\end{equation}
\end{small}

\vspace*{-5.5mm}
\noindent
See \cite[proof of Theorem 3]{yzl2016cdc} for additional details.

Now, we show that, when a MJLS is mean square
stabilizable, the optimal $T$-horizon cost function 
$\mathcal{J}_{_{\!T\!}}\!\left(\phi\right)$ corresponding to the solution to 
CDREs \eqref{eq:X_i_k} converges exponentially fast to the optimal 
infinite-horizon cost function $\hat{\mathcal{J}}_{_{\!\infty\!}}\!\left(\phi\right)$ related 
to the stabilizing solution to CAREs \eqref{eq:X_i}.

The next lemma provides a bound on the optimal cost of finite-horizon robust 
control, when the MJLS is stabilizible, and is the key in proving the 
convergence result.

\begin{lemma}\label{lemma:cost_boundedness}
Suppose that MJLS \eqref{eq:model_mjls_control} with PTI TPM \eqref{eq:polytope} 
is mean square stabilizable and detectable. 
Then, there exists $k'\!\in\!\mathbb{Z}_0$, such 
that for any $T\!\geq\!k'$, we have that the optimal $T$-horizon cost 
$\mathcal{J}_{_{\!T\!}}(\phi)$ of robust control can be bounded as

\vspace*{-2mm}
\begin{small}
\begin{equation}\label{eq:emss_bound_on_j_k}
\mathcal{J}_{_{\!T\!}}(\phi)\!\leq\!
\mathbb{E}\!\left(\left\|\hat{X}_{\theta_0}^{(\hat{v}_0)}\right\|\!+\!
\beta\zeta^{T}\!\left\|Z_{\theta_{T}}\!\right\|\!\right)\!
\left\|\mathrm{x}_0\right\|^2,
\end{equation}
\end{small}

\vspace*{-5mm}
\noindent
where $\hat{X}_{\theta_0}^{(\hat{v}_0)}$ is given by \eqref{eq:X_i}, $\hat{v}_0$ 
by \eqref{eq:hat_v_k}, $\beta$ and $\zeta$ by \eqref{eq:emss_bound_on_x}.

\vspace*{1mm}
\begin{proof}
By hypothesis, the considered MJLS is mean square 
stabilizable and detectable. So, the stabilizing solution $\{\hat{\bm X}_l\}_{l=1}^V$
for the control CAREs \eqref{eq:X_i} exists and is unique. Thus,
following the line of reasoning of \cite[Lemma 4]{zhang2009value}, consider the 
cost $\hat{\mathcal{J}}_{_{\!T\!}}(\phi)$ of robust control related to the trajectory 
$\left(\hat{\mathrm{x}}_k\right)_{k=1}^{T}$ generated from $\mathrm{x}_0$ by
application of the infinite-horizon stabilizing optimal control law  
$\left( \hat{\mathrm{u}}_k \right)_{k=0}^{T-1}$, with 
$\hat{\mathrm{u}}_k\!=\!\hat{K}_{\theta_k}^{(\hat{v}_k)}\mathrm{x}_k$ obtained
via \eqref{eq:r_i}\,--\eqref{eq:X_i}. By Bellman's principle of optimality,
stating that any segment of an optimal trajectory must be the optimal trajectory 
joining the two end points of the segment, and taking into account 
Definition \ref{definition:control_care}, \eqref{eq:emss}, 
\eqref{eq:emss_bound_on_x}, linearity of the expected value, sub-multiplicative 
property of the matrix norm, and that 
$Z_i\!\succeq\!0$, $\hat{X}_i^{(l)}\!\succeq\!0$ $\forall i\!\in\!\mathbb{M}$, 
$\forall l\!\leq\!L\!\leq\!V$, it follows that
\begin{small}
\begin{align}\label{eq:emss_bound_on_hat_j_k}
\hat{\mathcal{J}}_{_{\!T\!}}(\phi) =& 
	\,\hat{\mathcal{J}}_{_{\!\infty\!}}(\phi) \!-\!
\hat{\mathcal{J}}_{_{\!\infty\!}}(\hat{\mathrm{x}}_{T},\theta_T)\!+\!\mathbb{E}\!\left(
\hat{\mathrm{x}}^{*}_{T}Z_{\theta_{T}}\hat{\mathrm{x}}_{T}\right) \nonumber \\
=&\,\mathbb{E}\!\!\left(\!
\left\|\!\left(\!\hat{X}_{\theta_0}^{(\hat{v}_0)}\!
	\right)^{\!\!\frac{1}{2}}\!\!\!\mathrm{x}_0\right\|^2 \!\!\!\!-\!\!
\left\|\!\left(\!\hat{X}_{\theta_{T}}^{(\hat{v}_{T})}\!
	\right)^{\!\!\frac{1}{2}}\!\!\hat{\mathrm{x}}_{T}\right\|^2 \!\!\!\!+\!
\left\|\left(Z_{\theta_{T}}\!\right)^{\!\frac{1}{2}}\!
	\hat{\mathrm{x}}_{T}\right\|^2 \!\right) \nonumber\\
\leq&\,\mathbb{E}\!\left(\left\|\hat{X}_{\theta_0}^{(\hat{v}_0)}\right\|\! 
\left\|\mathrm{x}_0\right\|^2 \!\!+\!
\left\|Z_{\theta_{T}}\right\|\!\left\|\hat{\mathrm{x}}_{T}\right\|^2 \!
\right) \nonumber\\
\leq&\,\mathbb{E}\!\left(\left\|\hat{X}_{\theta_0}^{(\hat{v}_0)}\right\|\!+\!
\beta\zeta^{T}\!\left\|Z_{\theta_{T}}\right\|\!\right)\!
\left\|\mathrm{x}_0\right\|^2 .
\end{align}
\end{small}

\vspace*{-3.5mm}
Since the minimal $T$-horizon cost for the worst possible sequence of the 
transition probabilities is $\mathcal{J}_{_{\!T\!}}(\phi)$, we have that 
$\mathcal{J}_{_{\!T\!}}(\phi)\!\leq\!\hat{\mathcal{J}}_{_{\!T\!}}(\phi)$, and by 
\eqref{eq:emss_bound_on_hat_j_k} the lemma is proved.
\end{proof}
\end{lemma}

The next theorem shows that the sequence 
$\left(\mathcal{J}_{_{\!T\!}}\!\left(\phi\right)\right)_{T=0}^{\infty}$ 
of the finite-horizon optimal robust control cost functions converges to the 
infinite-horizon cost function $\hat{\mathcal{J}}_{_{\!\infty\!}}\!\left(\phi\right)$.  

\begin{theorem}\label{theorem:optimal_cost_convergence}
Suppose that MJLS \eqref{eq:model_mjls_control} with PTI TPM \eqref{eq:polytope} 
is mean square stabilizable 
and detectable. Then, for any initial condition $\phi$,
one has that
$\lim\limits_{T\to\infty}\mathcal{J}_{_{\!T\!}}\!\left(\phi\right) \!=\! 
\hat{\mathcal{J}}_{_{\!\infty\!}}\!\left(\phi\right)$.

\vspace*{0.5mm}
\begin{proof}
By hypothesis, the considered MJLS is mean square 
stabilizable and detectable. So, the stabilizing solution $\{\hat{\bm X}_l\}_{l=1}^V$
for the control CAREs \eqref{eq:X_i} exists and is unique.
By Theorem \ref{theorem:care}, the robust control cost function 
$\hat{\mathcal{J}}_{_{\!\infty\!}}\!\left(\phi\right)$ associated to 
$\{\hat{\bm X}_l\}_{l=1}^V$
is the optimal infinite-horizon cost function $\forall \phi$, so 
$\hat{\mathcal{J}}_{_{\!\infty\!}}\!\left(\phi\right)$ is minimal cost for the worst 
possible sequence of the transition probabilities. 
Then, by Lemma \ref{lemma:cost_boundedness},
there exists $k'$ such that
$\mathcal{J}_{_{\!T\!}}(\phi)\!\leq\!\hat{\mathcal{J}}_{_{\!T\!}}(\phi)$, for all $T\!\geq\!k'$.
Finally, from \eqref{eq:emss_bound_on_hat_j_k} and \eqref{eq:emss_bound_on_x}, 
it follows that
$\lim\limits_{T\to\infty}\mathcal{J}_{_{\!T\!}}\!\left(\phi\right) \!=\! 
\hat{\mathcal{J}}_{_{\!\infty\!}}\!\left(\phi\right)$.
\end{proof}
\end{theorem}

\section{ILLUSTRATIVE EXAMPLE}\label{sec:example}
Consider a simple economic system based on Samuelson's multiplier-accelerator 
model, which can be described by the following equations \cite[p. 8]{zhang2016analysis}:

\vspace*{-6.5mm}
\begin{equation*}
\mathcal{C}_t = c\mathcal{Z}_{t-1},\quad
\mathcal{I}_t = w(\mathcal{Z}_{t-1} - \mathcal{Z}_{t-2}),\quad
\mathcal{Z}_t = \mathcal{C}_t + \mathcal{I}_t + \mathcal{G}_t,
\end{equation*}

\vspace*{-2mm}
\noindent where 
$\mathcal{C}_t$ is the consumption expenditure, 
$\mathcal{Z}_t$ is the national income,
$\mathcal{I}_t$ is the induced private investment,
$\mathcal{G}_t$ is the government expenditure,
$s$ is the marginal propensity to save, ${1}/{s}$ is the multiplier,
$c\!=\!1\!-\!s$ is the marginal propensity to consume, i.e., a slope of the consumption versus income curve,
$w$ is the accelerator coefficient, and $t$ is the subscript for the discrete time.
As can be seen, the model is highly aggregated, intended primarily for use as a theoretical tool 
rather than as a realistic representation of the economy. However, it has been widely used 
in the literature on MJLSs, see e.g.~\cite{costa2006discrete,zhang2016analysis} and references therein.
Notably, if one sets 
the past national incomes as system states, and
the government expenditure as control input, i.e.,
$x_k \!\triangleq\!\left[\mathcal{Z}_{t-2},\mathcal{Z}_{t-1}\right]^T$,
$u_k \!\triangleq\! \mathcal{G}_t$, the above system can be rewritten in the state-space form
as in \eqref{eq:model_mjls_control}.
This system has three modes of operation, namely 
1, ``norm", with $s$ (or $w$) in mid-range; 
2, ``boom", having $s$ in low range (or $w$ in high range); 
3, ``slump", where $s$ is in high range (or $w$ in low range).
As a rationale for this terminology, one expects the marginal propensity to save $s$ 
to decline in ``good" times and increase in the ``bad", 
while
the acceleration coefficient $w$ would be expected to exhibit opposite tendencies
\cite{blair1975feedback}.

The parameters for each of these modes of operation and stochastic matrices 
used to bound the TPM are borrowed from \cite[pp. 180 -- 181, Example 8.3]{costa2006discrete}:

\vspace*{-4mm}
\begin{small}
\begin{equation*}
A_1 \!=\!\! \begin{bmatrix} 0 & 1 \\ -2.2308 & 2.5462 \end{bmatrix}\!, \; 
A_2 \!=\!\! \begin{bmatrix} 0 & 1 \\ -38.9103 & 2.5462 \end{bmatrix}\!, \;
\end{equation*}
\vspace*{-1.5mm}
\begin{equation*}
A_3 \!=\!\! \begin{bmatrix} 0 & 1 \\ 4.6384 & -4.7455 \end{bmatrix}\!, \;
B_1 \!=\!\! \begin{bmatrix} 0 \\ 1 \end{bmatrix}\!, \;
B_2 \!=\!\! \begin{bmatrix} 0 \\ 1 \end{bmatrix}\!, \;
B_3 \!=\!\! \begin{bmatrix} 0 \\ 1 \end{bmatrix}\!,
\end{equation*}
\vspace*{-1.5mm}
\begin{equation*}
C_1 \!=\!\! \begin{bmatrix} 1.5049 & -1.0709 \\ -1.0709 & 1.6160 \\ 0 & 0 \end{bmatrix}\!, \;
C_2 \!=\!\! \begin{bmatrix} 10.2036 & -10.3952 \\ -10.3952 & 11.2819 \\ 0 & 0 \end{bmatrix}\!, \;
\end{equation*}
\vspace*{-1.5mm}
\begin{equation*}
C_3 \!=\!\! \begin{bmatrix} 1.7335 & -1.2255 \\ -1.2255 & 1.6639 \\ 0 & 0 \end{bmatrix}\!, \;
D_1 \!=\!\! \begin{bmatrix} 0 \\ 0 \\ 1.6125 \end{bmatrix}\!, \;
D_2 \!=\!\! \begin{bmatrix} 0 \\ 0 \\ 1.0794 \end{bmatrix}\!,
\end{equation*}
\end{small}
\vspace*{-1.5mm}
\begin{small}
\begin{equation*}
D_3 \!=\!\! \begin{bmatrix} 0 \\ 0 \\ 1.0540 \end{bmatrix}\!, \;
P_4 \!=\!\! \begin{bmatrix} 1 & 0 & 0 \\ 0 & 1 & 0 \\ 0 & 0 & 1 \end{bmatrix}\!, \;
P_3 \!=\!\! \begin{bmatrix} 0.50 & 0.25 & 0.25 \\ 0.20 & 0.50 & 0.30 \\ 0.30 & 0.30 & 0.40 \end{bmatrix}\!, \;
\end{equation*}
\vspace*{-1.5mm}
\begin{equation*}
P_2 \!=\!\! \begin{bmatrix} 0.83 & 0.09 & 0.08 \\ 0.46 & 0.39 & 0.15 \\ 0.42 & 0.02 & 0.56 \end{bmatrix}\!,
P_1 \!=\!\! \begin{bmatrix} 0.51 & 0.25 & 0.24 \\ 0.14 & 0.55 & 0.31 \\ 0.10 & 0.18 & 0.72 \end{bmatrix}\!.
\end{equation*}
\end{small}

\vspace*{-2.5mm}
\indent
An approximate value of the TPM is obtained from the historical data 
of the United States Department of Commerce \cite[pp. 8\,--\,9]{zhang2016analysis}, while
the considered polytopic bounds were proposed by Costa et al. \cite[pp. 180 -- 181, Example 8.3]{costa2006discrete}.

Let us consider the MJLS with TPM $P(k)\!\in\!\mathrm{conv}\!\left\{P_i\right\}_{i=1}^{4}$.
This MJLS is unstable, since the spectral radii associated to the matrices $\Lambda_i$ computed via 
\eqref{eq:second_moment_matrix} are 
$\rho(\Lambda_1)\!=\!31.652$, 
$\rho(\Lambda_2)\!=\!20.110$,
$\rho(\Lambda_3)\!=\!29.962$, and 
$\rho(\Lambda_4)\!=\!38.910$, and the joint spectral radius 
$\hat{\rho}\!\left(_{\mathbb{V}}{\bm \Lambda}\right)\!\geq\!\max \rho(\Lambda_v)$.

This system is mean square stabilizable and detectable, so the optimal 
infinite-horizon robust controller exists. Specifically, there are two possible 
mode-dependent state-feedback gains, computed from \eqref{eq:k_i} via LMI 
optimization, and taking into account \eqref{eq:parsimonious_inf}, 
which are the following:

\vspace*{-4.5mm}
\begin{small}
\begin{equation*}
\hat{\bm K}_{\!1\!}\!=\!\!
\begin{pmatrix}
\!\!\hat{K}_{1}^{(\!\!\:1\!\!\:)} \!\!=\![ -2.223,  2.400],\!\! \\
\!\!\hat{K}_{2}^{(\!\!\:1\!\!\:)} \!\!=\![-38.860,  2.345],\!\! \\
\!\!\hat{K}_{3}^{(\!\!\:1\!\!\:)} \!\!=\![  4.632, -4.890]\!\!
\end{pmatrix}\!\!,\,\hat{\bm K}_{\!2\!}\!=\!\!
\begin{pmatrix}
\!\!\hat{K}_{1}^{(\!\!\:2\!\!\:)} \!\!=\![ -1.921,  1.538],\!\! \\
\!\!\hat{K}_{2}^{(\!\!\:2\!\!\:)} \!\!=\![-38.889,  2.392],\!\! \\
\!\!\hat{K}_{3}^{(\!\!\:2\!\!\:)} \!\!=\![  4.511, -5.407]\!\!
\end{pmatrix}\!.
\end{equation*}
\end{small}

\vspace*{-2.5mm}
\indent Notably, the control gain $\hat{\bm K}_{\!1\!}$ is obtained 
in correspondence of the vertex $P_3$, while $\hat{\bm K}_{\!2\!}$
is related to the vertex $P_4$. Both these state-feedback controllers 
are stabilizing, since

\vspace*{-2.5mm}
\begin{small}
\begin{equation*}
\hat{\rho}(_{\mathbb{V}}\hat{\bm \Delta}^{\!(1)}) \!<\! 0.05077,~
\hat{\rho}(_{\mathbb{V}}\hat{\bm \Delta}^{\!(2)}) \!<\! 0.66739,
\end{equation*}
\end{small}

\vspace*{-6mm}
\noindent where the value of the joint spectral radius has been computed 
via the JSR toolbox \cite{vankeerberghen2014jsr}. 

Noticeably, since the TPM $P_4$ 
treats all operational modes independently, the related control gains $\hat{\bm K}_2$
coincide with those obtained by solving the classical discrete-time algebraic Riccati equations
for linear time-independent systems.

The actual choice of the control gain depends on 
the initial condition 
$\phi$, as by \eqref{eq:hat_v_k}.
For $\mathrm{x}_0\!=\![1,1]^{T}\!=\!{\bm x}_0$, and $\theta_0\!=\!1$, we have that 
the optimal cost
$\mathcal{J}_{_{\!\infty\!}}({\bm x}_0,1) \!=\!\max\{495.715,\,6.161\}$ 
of the robust control is obtained
from the first parsimonious solution to the CAREs. For $\theta_0\!=\!2$,
we have instead that the optimal cost
$\mathcal{J}_{_{\!\infty\!}}({\bm x}_0,2) \!=\!\max\{2519.877,\,3478.062\}$ 
is achieved with the second solution, while for $\theta_0\!=\!3$, we find that
$\mathcal{J}_{_{\!\infty\!}}({\bm x}_0,3) \!=\!\max\{591.376,\,3.062\}$, so the
control gain to apply is $\hat{K}_{3}^{(\!\!\:1\!\!\:)}$.

Since the values of the joint spectral radius are relatively small, we have a 
fast convergence of the finite-horizon solution to steady-state. Let the terminal 
cost weighting matrices be $Z_1\!=\!2I_2$, $Z_2\!=\!I_2$, and $Z_3\!=\!4I_2$.
Then, for time horizon $T\!=\!8$ we have already almost the same values of the 
state-feedback gains, precisely

\vspace*{-4.5mm}
\begin{small}
\begin{equation*}
{\bm K}_{\!1\!}\!=\!\!
\begin{pmatrix}
\!\!{K}_{1}^{(\!\!\:1\!\!\:)} \!\!=\![ -2.223,  2.399],\!\! \\
\!\!{K}_{2}^{(\!\!\:1\!\!\:)} \!\!=\![-38.860,  2.344],\!\! \\
\!\!{K}_{3}^{(\!\!\:1\!\!\:)} \!\!=\![  4.632, -4.891]\!\!
\end{pmatrix}\!\!,\,{\bm K}_{\!2\!}\!=\!\!
\begin{pmatrix}
\!\!{K}_{1}^{(\!\!\:2\!\!\:)} \!\!=\![ -1.921,  1.538],\!\! \\
\!\!{K}_{2}^{(\!\!\:2\!\!\:)} \!\!=\![-38.889,  2.392],\!\! \\
\!\!{K}_{3}^{(\!\!\:2\!\!\:)} \!\!=\![  4.512, -5.403]\!\!
\end{pmatrix}\!,
\end{equation*}
\end{small}

\vspace*{-2.5mm}
\noindent while the optimal costs for the different initial operational modes are
respectively
$\mathcal{J}_{_{\!8\!}}({\bm x}_0,1) \!=\!\max\{495.698,\,6.160\}$,
$\mathcal{J}_{_{\!8\!}}({\bm x}_0,2) \!=\!\max\{2519.876,\,3478.062\}$, and
$\mathcal{J}_{_{\!8\!}}({\bm x}_0,3) \!=\!\max\{591.344,\,3.212\}$. So, it is evident that
for each initial operational mode the values of the finite-horizon optimal costs of 
the robust control are very close to the values of the optimal costs of the 
infinite-horizon setting. 

It is also worth noting that for any 
length $T$ of the finite-time horizon with the (integer) values in between $4$ and $1000$,
the maximum number of elements in the parsimonious set of CDREs is obtained for the time step 
$T\!-\!4$ and corresponds to $VL_{T-3}\!=\!4\cdot 4\!=\!16\!\ll\!4^{T}$.

If we consider the same problem for the MJLS with TPM 
$P(k)\!\in\!\mathrm{conv}\!\left\{P_i\right\}_{i=1}^{3}$, the illustrated approach
gives other two stabilizing control gains, first one associated to the vertex $P_1$,
and the second one, as before, related to the vertex $P_3$.

\vspace*{-4.5mm}
\begin{small}
\begin{equation*}
\tilde{\bm K}_{\!1\!}\!=\!\!
\begin{pmatrix}
\!\!\tilde{K}_{1}^{(\!\!\:1\!\!\:)} \!\!=\![ -2.222,  2.393],\!\! \\
\!\!\tilde{K}_{2}^{(\!\!\:1\!\!\:)} \!\!=\![-38.860,  2.331],\!\! \\
\!\!\tilde{K}_{3}^{(\!\!\:1\!\!\:)} \!\!=\![  4.629, -4.880]\!\!
\end{pmatrix}\!\!,\,\tilde{\bm K}_{\!2\!}\!=\!\!
\begin{pmatrix}
\!\!\tilde{K}_{1}^{(\!\!\:2\!\!\:)} \!\!=\![ -2.223,  2.400],\!\! \\
\!\!\tilde{K}_{2}^{(\!\!\:2\!\!\:)} \!\!=\![-38.860,  2.345],\!\! \\
\!\!\tilde{K}_{3}^{(\!\!\:2\!\!\:)} \!\!=\![  4.632, -4.890]\!\!
\end{pmatrix}\!,
\end{equation*}
\end{small}

\vspace*{-2.5mm}
\noindent with respective $\hat{\rho}(_{\mathbb{V}}\tilde{\bm \Delta}^{\!(1)}) \!<\! 0.03569$, 
$\hat{\rho}(_{\mathbb{V}}\tilde{\bm \Delta}^{\!(2)}) \!<\! 0.03610$.
The values of the joint spectral radius are particularly small
in this setting, so we expect a very fast convergence of the finite-horizon 
solution to steady-state. 

In fact, the optimal cost of the infinite-horizon robust state-feedback control for the
initial state ${\bm x}_0$ and different initial operational modes are
$\mathcal{J}_{_{\!\infty\!}}({\bm x}_0,1) \!=\!\max\{495.036,\,495.715\}$,
$\mathcal{J}_{_{\!\infty\!}}({\bm x}_0,2) \!=\!\max\{2613.443,\,2519.877\}$,
and
$\mathcal{J}_{_{\!\infty\!}}({\bm x}_0,3) \!=\!\max\{366.066,\,591.376\}$, while
for a finite-time horizon, with the length as short as $T \!=\! 5$, we have already 
exactly the same as for $\tilde{\bm K}_{\!1\!}$ and $\tilde{\bm K}_{\!2\!}$ values 
of the mode-dependent state-feedback gains (with precision up to the fourth decimal 
place). The related optimal costs of the robust control are 
$\mathcal{J}_{_{\!5\!}}({\bm x}_0,1) \!=\!\max\{495.021,\,495.715\}$,
$\mathcal{J}_{_{\!5\!}}({\bm x}_0,2) \!=\!\max\{2613.416,\,2519.853\}$,
and
$\mathcal{J}_{_{\!5\!}}({\bm x}_0,3) \!=\!\max\{366.051,\,591.358\}$. For
$T\!=\!6$ the finite- and infinite-horizon optimal costs coincide
(with the same precision, up to the fourth decimal place).

Noticeably, for any 
length $T$ of the finite-time horizon with the (integer) values in between $4$ and $1000$,
the maximum number of elements in the parsimonious set of CDREs is now obtained for the time step 
$T\!-\!2$ and corresponds to $VL_{T-1}\!=\!3\cdot 2\!=\!6\!\ll\!3^{T}$.

\section{CONCLUSIONS}\label{sec:conclusions}
In this paper, we generalized the infinite-horizon linear quadratic regulation results of 
stationary MJLSs to politopic time-inhomogeneous setting, providing an
analytical solution that can be computed efficiently. The natural extension of
this work is to consider different types of additive process and observation noise,
together with partial and delayed information on operational modes.

\enlargethispage{-10.8cm} 

%
\bibliographystyle{IEEEtran} 
\bibliography{cdc18bib}
\end{document}